\documentclass{emulateapj}
\slugcomment{Accepted: December 11, 2015}
\shorttitle{B-field amplification by shocks in partially ionized ISM}

\shortauthors{Yutaka Ohira}

\begin{document}

\title{Magnetic field amplification by collisionless shocks in partially ionized plasmas}

\author{Yutaka Ohira}

\begin{abstract}
In this paper, we study shock structures of collisionless shocks in partially ionized plasmas by means of two-dimensional hybrid simulations, where the shock is a perpendicular shock with shock velocity $v_{\rm sh}\approx 40~v_{\rm A} \approx 1333~{\rm km/s}$ and the upstream ionization fraction is 0.5. 
We find that large density fluctuations and large magnetic fields fluctuations are generated both in the upstream and downstream regions. 
In addition, we find that the velocity distribution of downstream hydrogen atoms has three components. 
Observed shock structures suggest that diffusive shock acceleration can operate at perpendicular shocks propagating into partially ionized plasmas in real three-dimensional systems.  
\end{abstract}


\affil{Department of Physics and Mathematics, Aoyama Gakuin University, 
5-10-1 Fuchinobe, Sagamihara 252-5258, Japan; ohira@phys.aoyama.ac.jp}
\section{Introduction}
Collisionless shocks driven by supernova explosions have been expected to accelerate cosmic rays (CRs). 
The acceleration mechanism is though to be diffusive shock acceleration (DSA) \citep{axford77,krymsky77,bell78,blandford78}. 
In fact, recent gamma-ray observations show that supernova remnants (SNRs) accelerate CRs \citep{koyama95,ohiraetal11}. 
Magnetic fields play a crucial role in the particle acceleration, but our understanding of magnetic fields around SNR shocks is still incomplete. 
Several observations suggest that magnetic fields are amplified over $100 \mu {\rm G}$ around SNR shocks \citep{vink03, berezhko03, bamba05, uchiyama07}. 
In addition, many theoretical studies also suggest that magnetic fields are amplified by several mechanisms around SNR shocks. 
However, what is the dominant mechanism remains an open question.

The interstellar medium and ejecta of SNRs are often a partially ionized plasma. 
Plasma instabilities are often thought to be stabilized in the partially ionized plasma because of a collision with neutral particles. 
However, it was recently proposed that many plasma instabilities are excited by ionization around a collisionless shock in the partially ionized plasma \citep{raymond08,ohiraetal09,ohira10,ohira14}. 
\citet{ohira13} performed the first hybrid simulation of a collisionless shock wave 
propagating into a partially ionized plasma and showed that ionization of neutral particles excite plasma instabilities both in the upstream and downstream regions. 
In addition, the simulation showed that some downstream hydrogen atoms leak into the upstream region. 
In the upstream region (neutral precursor region), density fluctuations are excited by ionization of leaking neutral particles from the downstream region \citep{ohira13,ohira14}. 
According to early studies, upstream density fluctuations amplify magnetic fields in the downstream region if the Alfv{\'e}n Mach number and amplitude of the density fluctuations are sufficiently large \citep{giacalone07, inoue09}. 
For the first hybrid simulation of \citet{ohira13}, the upstream flow velocity is $v_{\rm d}=10~v_{\rm A}=2000~{\rm km/s}$ in the downstream rest frame, where $v_{\rm A}$ is the Alfv{\'e}n velocity. 
The amplitude of the upstream density fluctuation is about $\delta \rho/\rho_0 \approx 0.5$, 
so that the expected velocity dispersion of the downstream turbulence is not more than about $0.25 ~v_{\rm d}$, 
that is, the kinetic energy of turbulence in the downstream region is not more than $6.25~\%$ of the total shock kinetic energy. 
If all the kinetic energy of turbulence was converted to the magnetic field energy,  
the expected magnetic field would be at most $2.5~B_0$, which is smaller than the shock-compressed magnetic field, $4~B_0$, where $B_0$ is the upstream magnetic field strength. 
Therefore, upstream density fluctuations cannot stretch compressed magnetic field lines in the downstream region for \citet{ohira13}.

More leakage of neutral particles from the downstream region and larger density fluctuations can be expected for a slower shock velocity than that of \citet{ohira13} 
because the charge exchange rate becomes larger than the collisional ionization rate \citep{blasi12,ohira12}. 
In this paper, we perform two-dimensional hybrid simulations of collisionless shocks propagating into the partially ionized plasmas for $v_{\rm d}= 30~v_{\rm A} = 1000~{\rm km/s}$, that are higher Alfv{\'e}n Mach number and slower shock velocity than that of the first simulation. 
We then show that the shock strongly amplify magnetic fields.
\section{Hybrid Simulations}
\label{sec:2} 
\begin{figure*}\centering
\includegraphics[width=0.85\textwidth]{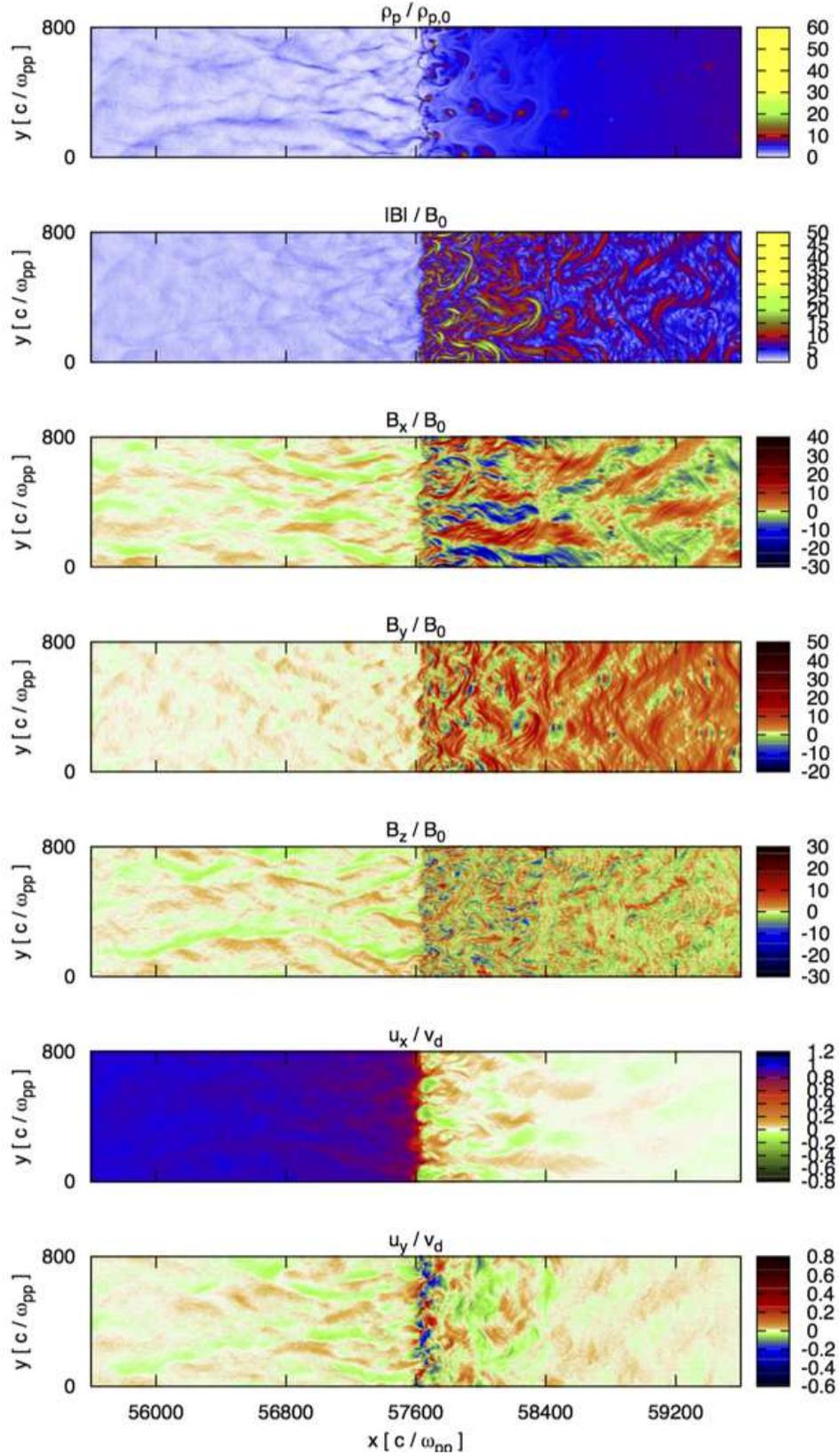}
\caption{Two-dimensional shock structures at $t=2000~\Omega_{\rm cp}^{-1}$. From top to bottom, the panels show proton density, $\rho_{\rm p}$, magnetic field strength, $|B|$, each component of magnetic field, $B_{\rm x},~B_{\rm y},~B_{\rm z}$, and fluid velocity of protons in the downstream rest frame, $u_{\rm x},~u_{\rm y}$. The left regions ($x<57600~c/\omega_{\rm pp}$) are upstream regions.
\label{fig:1}}
\end{figure*}

In order to study collisionless shocks propagating into partially ionized plasmas, 
we use a two-dimensional hybrid code that computes the motion of protons and hydrogen atoms, Maxwell's equations in the low-frequency limit, and ionization of hydrogen atoms \citep{ohira13}. 
The code solves charge exchange and collisional ionization as ionization of hydrogen atoms. 
Their cross sections are obtained from \citet{barnett90} and \citet{janev93}. 
The ionization timescale is at least $10^5$ times longer than the gyro period for typical young SNRs.
To solve this large timescale gap exactly is challenging even for current supercomputers. 
Therefore, we adopt cross sections boosted by a factor of $10^3$, but ionization timescale is still much longer than the gyro period.

Because hybrid codes do not solve the electron dynamics, the evolution of the electron temperature must be assumed. 
Although we have not understood how electrons are heated in collisionless shocks, 
the electron temperature is expected to be $T_{\rm e} \approx 0.01 m_{\rm p}v_{\rm d}^2/3$ in the downstream region of SNRs \citep{ghavamian07,ohira07,ohira08,rakowski08,van08,laming14,vink15}, 
where $m_{\rm p}v_{\rm d}^2/3$ is the downstream proton temperature for the strong shock limit and $v_{\rm d}$ is the upstream flow velocity in the downstream rest frame. 
Therefore, when we calculate collisional ionization by electrons, we assume that the relative velocity between a hydrogen atom and electrons is given by \citep{pauls95}
\begin{equation}
v_{\rm rel,eH} = \sqrt{\frac{8kT_{\rm e}}{\pi m_e}+|\mbox{\boldmath $v$}_{\rm H}-\mbox{\boldmath $u$}_{\rm p}|^2}~~,
\end{equation}
where $\mbox{\boldmath $v$}_{\rm H}$ is the velocity of the hydrogen atom and the mean velocity of electrons is assumed to be the mean velocity of protons, $\mbox{\boldmath $u$}_{\rm p}$, and the electron temperature is assumed to be $T_{\rm e} = 0$ and $T_{\rm e}=0.01m_{\rm p}v_{\rm d}^2/3$ in the upstream and downstream regions, respectively.

\subsection{Simulation Setup}
We perform two-dimensional hybrid simulations in the $xy$ plane. 
We inject simulation particles with a positive drift velocity at the left boundary, $x=0$, and 
the simulation particles are specularly reflected at the right boundary. 
Then, a collisionless shock is produced and propagates to the left boundary as time goes on, that is, the simulation frame corresponds to the downstream rest frame. 
We impose the periodic boundary condition in the $y$ direction. 
The simulation box size is $L_{\rm x} \times L_{\rm y} = 77600~c / \omega_{\rm pp} \times 800~c / \omega_{\rm pp}$, where $c$ and $\omega_{\rm pp}$ 
are the speed of light and plasma frequency of protons, respectively. 
The cell size and time step are 
$\Delta x = \Delta y = c / \omega_{\rm pp}$ and $\Delta t=8.333\times 10^{-3}~\Omega_{\rm cp}^{-1}$, respectively.
Initially, we put 64 protons and 64 hydrogen atoms in each cell and 
the magnetic field is taken to be spatially homogenous, pointing 
in the $y$ direction, $\mbox{\boldmath $B$}=B_0 \mbox{\boldmath $e$}_{\rm y}$. 
Therefore, a perpendicular shock is formed in this simulation, where the shock normal direction is perpendicular to upstream mean magnetic field lines. 
For parameters of the upstream plasma, we set $f_{\rm i}=0.5,\beta_{\rm p}=\beta_{\rm H}=0.5$, and $v_{\rm d}=30~v_{\rm A}=1000~{\rm km/s}$, where $f_{\rm i},\beta_{\rm p},\beta_{\rm H}$, and $v_{\rm d}$ are the upstream ionization fraction, ratio of the proton pressure to the magnetic pressure, ratio of the hydrogen pressure to the magnetic pressure, and drift velocity of the $x$ direction, respectively. 
$v_{\rm A}=B_0/\sqrt{4\pi \rho_{\rm p,0}}$ is the Alfv{\'e}n velocity defined by the proton mass density in the far upstream region, $\rho_{\rm p,0}$.

\subsection{simulation results}
\begin{figure}
\includegraphics[scale=0.72]{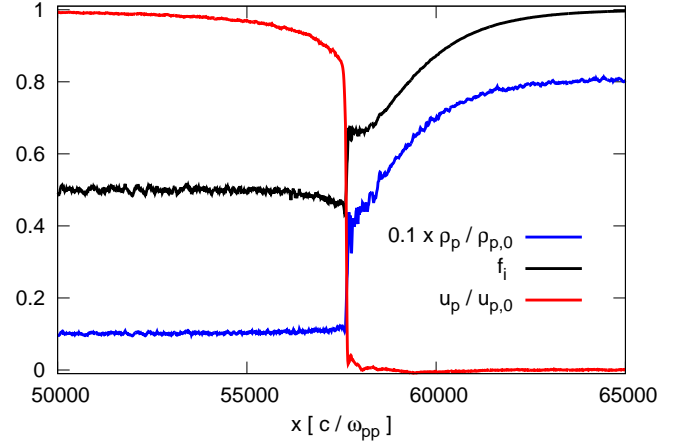}
\caption{Shock structures averaged over the $y$ direction at $t=2000~\Omega_{\rm cp}^{-1}$. The red, blue, and black lines
show the fluid velocity normalized by the far upstream value, $u_{\rm x}/v_{\rm d}$, the proton density normalized by 10 times the far upstream value, $0.1\rho_{\rm p}/\rho_{\rm p,0}$, and the ionization fraction, $f_{\rm i}$, respectively. 
\label{fig:2}}
\end{figure}
\begin{figure}
\includegraphics[scale=0.75]{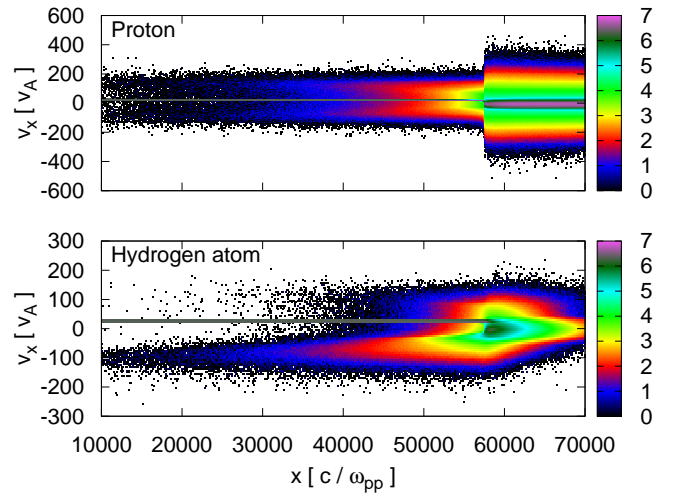}
\caption{Phase space plots of protons (top) and hydrogen atoms (bottom) at$t=2000~\Omega_{\rm cp}^{-1}$. The color shows the
phase space density in logarithmic scale. 
\label{fig:3}}
\end{figure}
\begin{figure}
\includegraphics[scale=0.7]{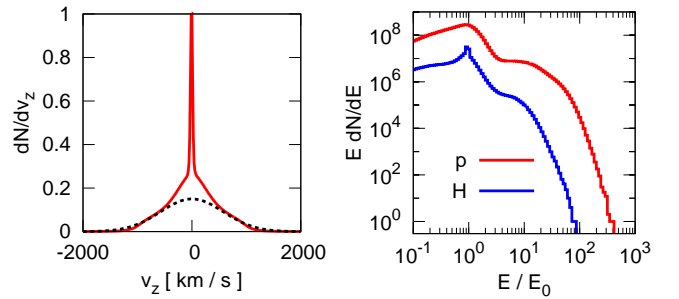}
\caption{Velocity distribution of hydrogen atoms (left) and energy spectra (right) in the downstream region, $57700~c/\omega_{\rm pp}\le x \le 72900~c/\omega_{\rm pp}$,$t=2000~\Omega_{\rm cp}^{-1}$. 
\label{fig:4}}
\end{figure}

In Figure~\ref{fig:1}, we show shock structures of the proton density, $\rho_{\rm p}/\rho_{\rm p,0}$, magnetic field $|B|/B_0,~B_{\rm x}/B_0,~B_{\rm y}/B_0,~B_{\rm z}/B_0$, and fluid velocity in the downstream rest frame, $u_{\rm x}/v_{\rm d},~u_{\rm y}/v_{\rm d}$, at time $t=2000~{\Omega_{\rm cp}}^{-1}$, 
where $u_{\rm x}$ and $u_{\rm y}$ are x and y components of the mean proton velocity, respectively. 
The shock front is located at about $x=57600~c/\omega_{\rm pp}$ and 
the left side of the shock front is the upstream region.
The shock velocity is $v_{\rm sh} \approx 4v_{\rm d}/3 = 40~v_{\rm A}=1333~{\rm km/s}$ in the upstream rest frame.  
If we use another Alfv{\'e}n velocity, $v_{\rm A,tot}=B_0/\sqrt{4\pi(\rho_{\rm p,0}+\rho_{\rm H,0})}$, defined by the total mass density, the shock velocity is expressed as $v_{\rm sh} \approx 57~v_{\rm A,tot}$ in the upstream rest frame, where $\rho_{\rm H,0}$ is the upstream hydrogen mass density.

Large amplitude density fluctuations are observed both in the upstream and downstream regions. 
In our previous simulation for $v_{\rm d} = 10~v_{A}=2000~{\rm km/s}$, 
the proton density is correlated with the magnetic field strength in the upstream region and the wave vector is highly oblique to the magnetic field direction, that is, the fast magnetosonic mode are excited \citep{ohira14}.  
In this simulation for $v_{\rm d} = 30~v_{A}=1000~{\rm km/s}$, the correlation is observed in many upstream regions, 
but unticorrelation between the proton density and the magnetic field strength is observed in high density filaments of the upstream region and the wave vector is almost parallel to the initial magnetic field direction. 
The velocity field in Figure~\ref{fig:1} ($u_{\rm y}$) show that the upstream filamentary structures are produced by compression in the $y$ direction. 
The compression motion is driven by the magnetic pressure of large amplitude Alfv{\'e}n waves propagating to the initial magnetic field direction (see $B_{\rm x}$ and $B_{\rm z}$ in Figure~\ref{fig:1}). 
The upstream Alfv{\'e}n waves are excited by the pressure anisotropy produced by ionization of leaking neutral particles \citep{ohira13}.

The amplitude of the filamentary structures in the upstream region is about $\delta \rho_{\rm p}/\rho_{\rm p,0} \approx 5$. 
Such large density fluctuations produce rippled shock structures and turbulence in the downstream region (see velocity structures $u_{\rm x}$ and $u_{\rm y}$ in Figure~\ref{fig:1}). 
As the result, downstream magnetic fields are amplified by the turbulence, which was suggested by \citet{ohira13}. 
In this paper, we first demonstrated via hybrid simulations that even though shocks propagate partially ionized plasmas, magnetic fields are amplified up to near the equipartition level by the downstream turbulence. 
Figure~\ref{fig:1} shows that downstream magnetic fields are locally amplified over $30B_0$ and the spatially averaged value is about $15 B_0$ at $x = 58400~c / \omega_{\rm pp}$, that is, significant fraction of the upstream kinetic energy is converted to the magnetic field energy. 
After the magnetic field strength reaches a saturation level, it gradually decays as the turbulence decays. 
The decay length scale is about $300~c/{\omega_{\rm pp}}$ which is much smaller than the observable length scale. 
However, once particles are accelerated by DSA, density fluctuations with larger length scale would be generated by the accelerated particles \citep{caprioli13}. 
If so, the decay length scale of magnetic fields would be larger. 
Although simulations in this paper are two-dimensional system, it has been demonstrated by the three-dimensional magnet hydrodynamic simulation that the turbulence do amplify magnetic fields up to near the equipartition level \citep{inoue10}. 
Therefore, actual higher Alfv{\'e}n Mach number shocks can be expected to amplify magnetic fields more strongly.

The amplitude of the magnetic field fluctuations is $\delta B/B_0\approx 3$ in the upstream region. 
The coherent length scale of magnetic field structures is of the order of the gyroradius of upstream pickup ions. 
The pickup ions are preferentially accelerated by charge exchange and pickup processes \citep{ohira13}. 
Therefore, the observed magnetic turbulence is important for injection into DSA. 
Simulations in this paper are performed in the two-dimensional system, 
so that particle diffusion perpendicular to the magnetic field line is artificially surpassed \citep{jokipii93,giacalone94}. 
Therefore, we can expect DSA in perpendicular shocks propagating into partially ionized plasmas for a real three dimensional system, which will be addressed in future works.

There are no clear mirror mode structures in the downstream region although we observed that in the previous simulation for $v_{\rm d} = 10~v_{A}=2000~{\rm km/s}$. 
The mirror mode is excited by the pressure anisotropy of pickup ions produced in the downstream region \citep{raymond08}. 
However, magnetic field lines are not ordered in the downstream region of this simulation, 
so that pickup ions produced in the downstream regions do not have pressure anisotropy and the mirror mode is not excited.

In Figure~\ref{fig:2}, we show one-dimensional shock structures averaged over the $y$ direction at time $t=2000~\Omega_{\rm cp}^{-1}$. 
The red, blue, and black lines indicate the mean proton velocity of the $x$ direction, $u_{\rm p}/u_{\rm p,0}$, the proton density, $0.1\rho_{\rm p}/\rho_{\rm p,0}$, and ionization fraction, $f_{\rm i}$, respectively.
As with our previous simulation, the plasma flow is gradually decelerated within the ionization length scale but the shock modification is larger than that of the previous simulation for $v_{\rm d}=2000~{\rm km/s}$. 
This is because more hydrogen atoms leak into the upstream region for $v_{\rm d}=1000~{\rm km/s}$ than that for $v_{\rm d}=2000~{\rm km/s}$. 
The number of leaking neutral particles increases with the ratio of the charge exchange rate to the collisional ionization rate.
The ratio decreases with the shock velocity, so that many hydrogen atoms leak into the upstream region in this simulation compared with our previous simulation. 
The ratio of the fluid velocity just front of the shock to that just behind the shock is about 3.2 and the total compression ratio is about 4 in this simulation. 
If particles are accelerated by DSA in this velocity structure, the momentum spectrum, $dN/dp$, becomes steeper than $p^{-2}$ \citep{ohira12,blasi12}.

Figure~\ref{fig:3} shows the phase space at time $t=2000~{\Omega_{\rm cp}}^{-1}$. 
As one can see, there are many leaking hydrogen atoms in the upstream region. 
Downstream hot hydrogen atoms produced by charge exchange can leak into the upstream region because they do not interact with electromagnetic fields and some of them have a velocity faster than the shock velocity. 
Because the ionization rate dominates over the charge exchange rate for $v_{\rm rel}>3000~{\rm km/s}$, 
the maximum velocity of leaking neutral particles is about $3000~{\rm km/s}$ that is $90~v_{\rm A}$ in this simulation.
The leaking hydrogen atoms are ionized by upstream particles and picked up by the upstream flow and become pickup ions. 
All the upstream protons are mainly thermalized at the collisionless shock at $x=57600~c/\omega_{\rm pp}$.

In the left panel of Figure~\ref{fig:4}, we show the z component of the velocity distribution of hydrogen atoms in the downstream region, 
$57700~c/\omega_{\rm pp} \leq x \leq72900~c/\omega_{\rm pp}$, 
at time $t=2000~{\Omega_{\rm cp}}^{-1}$. 
For the standard model of H$\alpha$ emission from SNRs \citep{chevalier78}, 
the velocity distribution has two components (narrow and broad).
The dashed line in the left panel of Figure~\ref{fig:4} shows a Gaussian distribution with dispersion $\sigma^2=2v_{\rm d}^2/3$, that corresponds to the broad component of the standard model.
As one can see, the velocity distribution has three components (narrow, intermediate and broad) in this simulation. 
The intermediate component originates from protons thermalized by the collisionless shock. 
Because upstream cold protons are significantly decelerated before they interact with the collisionless shock (see Figure~\ref{fig:2}), the downstream temperature of these protons becomes lower than that of the standard model. 
The lost kinetic energy of upstream cold protons mainly converts to the kinetic energy of pickup ions. 
As a result, the width of intermediate component becomes smaller than expected from the standard model. 
The broadest component is produced by protons ionized in the downstream region. 
Upstream neutral particles penetrate the collisionless shock front without energy exchange with pickup ions, 
so that protons ionized in the downstream region have the kinetic energy of $m_{\rm p}v_{\rm d}^2/2$. 
Therefore, the width of the broadest component is the same as that of the standard model. 
As with the standard model, the narrowest component originates from upstream hydrogen atoms (before charge exchange). 
We do not see any heating of upstream hydrogen atoms in this simulation, so that 
the velocity dispersion of the narrowest component in this simulation corresponds to the temperature of the far upstream region. 
\citet{morlino12} proposed the three-component structure of H$\alpha$ emission, but the formation mechanism is different from our results. 
Their intermediate component originates from the neutral precursor region, 
and the velocity dispersion is of the order of $100~{\rm km/s}$, which is much smaller than that in this simulation. 
Interestingly, the three-component structure is observed in the SNR Tycho \citep{raymond10}.
If the broadest component cannot be identified as H$\alpha$ emission, we would mistakenly identify the intermediate component as the broad competent of the standard model of H$\alpha$ emission from SNRs and mistakenly estimate the downstream temperature. 
Therefore, estimating the downstream temperature from H$\alpha$ emission requires careful attention.

The right panel of Figure~\ref{fig:4} shows energy spectra of protons and hydrogen atoms in the downstream region, 
$57700~c/\omega_{\rm pp} \leq x \leq72900~c/\omega_{\rm pp}$, at time $t=2000~{\Omega_{\rm cp}}^{-1}$. 
Some protons are accelerated to over $10$ times the initial kinetic energy, $E_0=m_{\rm p} v_{\rm d}^2/2$. 
As with previous simulation, particles are accelerated by charge change and pickup processes \citep{ohira13}. 
The typical energy becomes $E=10 E_0$ after the first cycle of the acceleration \citep{ohira13}. 
Almost all accelerated particles experience only the first cycle of the acceleration and the energy spectrum has a cutoff at $E\approx 10E_0$ in the previous simulation for $v_{\rm d} = 2000~{\rm km/s}$. 
On the other hand, the energy spectrum has a cutoff at $E\approx 60E_0$ in this simulation for $v_{\rm d}=1000~{\rm km/s}$ because some accelerated particle can experience the second cycle of the acceleration.
The maximum velocity of leaking neutral particles is about $3000~{\rm km/s}$ in the downstream rest frame. 
After ionization in the upstream region, they are picked up by the upstream flow and their velocity becomes about $4000~{\rm km/s}$ in the upstream rest frame. The picked up particles interact with the shock and are heated to $v\approx 8000~{\rm km/s}$, that is, the kinetic energy becomes $E\approx 64~E_0$ that is comparable to the cutoff energy in this simulation. 
The total kinetic energy of non thermal particles is about $29~\%$ 
of the total kinetic energy of all particles in this simulation, where the nonthermal particles are defined by $E>4~E_0$. 

As mentioned above, particle diffusion perpendicular to the magnetic field line is artificially surpassed in our two-dimensional simulations, so that we cannot follow further acceleration by DSA. 
However, several early studies have already show that once CRs are injected, magnetic field fluctuations are excited by CRs even though there are neutral hydrogen atoms \citep{drury96,bykov05,reville07}. 
Therefore, we can expect the further acceleration by DSA in a real three dimensional system.

\section{Discussion}
We briefly mention simulation results for other shock velocities in order to discuss dependence on the upstream velocity (shock velocity) and Alfv{\'e}n Mach number. 
The upstream velocities of our previous simulation and this simulation are $v_{\rm d}=10~v_{\rm A}=2000~{\rm km/s}$ and $v_{\rm d}=30~v_{\rm A}=1000~{\rm km/s}$, respectively. 
In addition to that, we perform two other simulations for $v_{\rm d}=10~v_{\rm A}=1000~{\rm km/s}$ and $v_{\rm d}=30~v_{\rm A}=2000~{\rm km/s}$. 
For $v_{\rm d}=10~v_{\rm A}=1000~{\rm km/s}$, phase space plot, one-dimensional shock structures, velocity distribution and energy spectra are similar to that of this simulation for $v_{\rm d}=30~v_{\rm A}=1000~{\rm km/s}$, but the large magnetic field amplification is not observed. 
For $v_{\rm d}=30~v_{\rm A}=2000~{\rm km/s}$, phase space plot, one-dimensional shock structures, velocity distribution and energy spectra are similar to that of our previous simulation for $v_{\rm d}=10~v_{\rm A}=2000~{\rm km/s}$, and magnetic fields are slightly amplified. 
Therefore, only magnetic fields strongly depend on the Alfv{\'e}n Mach number of shocks and other strongly depend on the shock velocity. 
All results should be depend on other parameters (ionization fraction, electron temperature, and magnetic field orientation). 
Systematic studies will be address in future works. 

In this paper, we use artificially large cross sections of charge exchange and collisional ionization in order to reduce the ionization length scale and time scale. 
For actual SNR shocks, the length scale of neutral precursor region becomes larger than that of our simulations.  
Then, the size of turbulent regions would be lager than that of this simulation, and the second order acceleration would become important \citep{ohira13a}. 
The growth rates of instabilities in this simulation might be artificially increased by a factor of $10^3$. 
In actual SNR shocks, amplitudes of excited waves would be smaller than that in this simulation 
because of several damping mechanisms. 
On the other hand, the Alfv{\'e}n Mach number of actual young SNRs is larger than that in this simulation, 
that is, the free energy of leaking particles of realistic young SNRs is larger than that in this simulation. 
Therefore, amplitudes of density and magnetic field fluctuations would be smaller or larger than that in this 
simulation. 
In order to understand more precisely, we need to perform a more realistic simulation. 
\section{Summary}
In this paper, we have performed two-dimensional hybrid simulations in order to investigate collisionless shocks generated by SNRs in partially ionized plasmas. 
We have found that large magnetic field and density fluctuations are excited both in the upstream and downstream regions for high Alfv{\'e}n Mach number shocks. 
In addition, we have found that the velocity distribution of downstream hydrogen atoms has three components for $v_{\rm sh} \approx 4v_{\rm d}/3=1333~{\rm km/s}$. 
As with our previous simulation, we have observed leaking neutral particles, the modified shock structure, and particle acceleration by charge exchange and pickup processes in this simulations. 
These results suggest that collisionless perpendicular shocks propagating into partially ionized plasma can produce cosmic rays. 
\acknowledgments
We thank T. Inoue and R. Yamazaki for useful comments. 
We are also grateful to ISSI (Bern) for support of the team "Physics of the injection of particle acceleration at  astrophysical, heliospheric, and laboratory collisionless shocks". 
Numerical computations were carried out on the XC30
system at the Center for Computational Astrophysics (CfCA)
of the National Astronomical Observatory of Japan.

\end{document}